# Extraction Urban Clusters from Geospatial Data: A Case Study from Switzerland

Jingya YAN[1]

**Abstract**

Different techniques were developed to extract urban agglomerations from a big dataset. The urban agglomerations are used to understand structure and growth of cities. However, the major challenge is to extract urban agglomerations from the big data, which can reflect human activities. Community urban cluster refers to spatially clustered geographic events, such as human settlements or activities. It provides a powerful and innovative insight to analyze the structure and growth of the real city. In order to understand the shape and growth of urban agglomerations in Switzerland from spatial and temporal aspects, this work identifies urban clusters from nighttime light data and street network data. Nighttime light data record lights emitted from human settlements at night on the earth's surface. This work uses DMSP-OLS Nighttime light data to extract urban clusters from 1992 to 2013. The street is one of the most important factors to reflect human activities. Hence, urban clusters are also extracted from street network data to understand the structure of cities. Both of these data have a heavy-tailed distribution, which includes power laws as well as lognormal and exponential distributions. The head/tail breaks is a classification method to find the hierarchy of data with a heavytailed distribution. This work uses head/tail breaks classification to extract urban clusters of Switzerland. At last, the power law distribution of all the urban clusters was detected at the country level.

[1] ETH Zurich, Future Resilient Systems, Singapore-ETH Centre
Corresponding author:
Jingya YAN, ETH Zurich, Future Resilient Systems, Singapore-ETH Centre, 1 Create Way, #06-01 CREATE Tower, Singapore 138602.
Email: jingya.yan@frs.ethz.ch

# Introduction

Urbanization is a relatively new global issue. Rapid urbanization and accelerating socio-economic development have generated global problems from climate change and its environmental impacts to incipient crises in food, energy and water availability, public health, financial markets and the global economy (Bettencourt and West2010). The worldwide rapid urbanization presents an urgent challenge for developing a predictive, quantitative theory of urban organization and sustainable development. One of the most important factors of city growth is population. As recently as 1950, only 30% of the world's population lived in the urban areas. Today, more than half live in urban centers, among which the developed countries have about 80% urban. Almost 2 billion people will move to cities around 2050, especially in China, India and other developing countries (Moreno et al.2010). As the main body of the city, human activities influence the development of city infrastructure directly. In order to learn the growth and evolution of the city, it is necessary to analyze the changes of the urban area and the relationship with human behaviors.

As a kind of big data, spatial data is widely used in urban studies and spatial analysis. On the one hand, remote sensing data is the most common data to delineate the cities' boundaries. The nighttime light (NTL) data is a kind of remote sensing data to detect the lights from human settlement, gas flares and fires, which is derived from the Defense Meteorological Satellite Programs Operational Linescan System (DMSP/OLS). In recent years, NTL data become popular in urban studies or related research (e.g. Doll(2008) and Elvidge et al.(2009)). On the other hand, the street is a fundamental unit to consist of a city, which reflects human settlement and development of transportation system directly. The street line is a kind of basic spatial vector data. Both of these data are useful to analyze human settlements and growth of urban areas at global and regional scale.

How to identify urban areas from a large set of spatial data? Some previous works to realize the delineation of urban area are based on the administrative boundary. This is a top-down method to observe a city, which relies on census data in a subjective area. The traditional definition of a city relies on its administrative boundary, whereas the urban sprawl and growth process resulted in urban clusters that are crossing the traditional geographic delineations of cities. This phenomenon raises the question how geographic urban agglomerations can be delineated for a specific point in time. The boundaries of the urban clusters are not as same as the administrative boundaries of the corresponding cities. The urban clusters provide a bottom-up thinking to define a collective region includes a cluster of fundamental units (e.g. street blocks), which is also called natural cities (Jiang and Jia2011). The extraction of urban clusters relies on the structure of data without any subjective views of human influence. The urban cluster is a powerful concept for studying urban structure and dynamics with a new insight into the evolution of real cities.

This work aims to extract urban clusters from spatial data in order to study the shape and growth and of urban agglomerations in Switzerland. Compared with some large regions at the global level, the area of Switzerland and its cities are small. The urban clusters from NTL data are used to understand the change of urban areas in a time series. Meanwhile, the street network data provides much more information about human settlement to support the research on the structure of real cities. Overall, this work should provide a different way of thinking about the study city growth in the small region scale.

This paper is organized as follows: the second section introduces the existing work of urban study through spatial data. In addition, the head/tail breaks classification is reviewed. The third section introduces the extraction of urban clusters from two kinds of data, includes the basic information of data and extraction process of urban clusters. The fourth section represents the results and discussion of this work. The last section presents conclusions and direction for future works on extending the extraction of urban clusters to understand shape and growth of the city.

## Background

### *Urban Area Detection from Spatial Data*

Related to different purposes, a large amount of studies work on extraction of urban area or from various kinds of data. For instance, social media data is used to find the social groups for analysis of structure (Girvan and Newman2002) and social network (Papadopoulos et al.2012). Street network, as a kind of spatial data, is one of the most important factors to reflect human activities (Jiang and Jia2011). Hence, it is widely used to analyze urban morphology. Such as, Buhl et al.(2006) analyzed human settlement from topological pattern of street network and Masucci et al.(2013) used street network to analyzed urban growth. Most of the existing works focus on the analysis of topological, logical, and structural properties in the street network.

Nighttime lights data is a kind of remote sensing data that records city lights, fires, gas flares and so on. Nighttime lights provide a versatile and user friendly data source for the social scientist, whether it is used simply to define an urban area or used more intensively to model population, economic activity or some other socio-economic parameter (Doll2008). However, the initial objective of night time light data was not observe human activities. In recently thirty years, NTL data began to be used for analysis of human settlement and activities. Croft(1978) was the first time to identify the potential of NTL data as an indicator of human activity. Since then, studies have shown strong relationships between NTL data and key socioeconomic variables. For instance, Sutton et al.(2007) correlated NTL data and population density to estimates of gross domestic product (GDP) at subnational level. Elvidge et al.(1997b);Sutton (1997);Amaral et al.(2006) used NTL data to detect human settlements and estimate the urban population. In addition, NTL data is popularly used to analyze urban areas. Elvidge et al.(1997a);Kasimu et al.(2009) developed NTL data processing methods which aim to map city lights or characterize urban features. All of these works are static analysis for an individual time of NTL data. Multi-temporal NTL data have been used to map exurban change with a focus on new urban development into the fire prone zone (Cova et al.2004) and urbanization dynamics at regional and global scales (Zhang and Seto2011).

Even through there are some common approaches to classify the spatial data and extract urban areas, they are not appropriate for non-linear datasets, such as quantile. Figure d1 represents an example of urban areas extraction from NTL data uses quantile. On the one hand, comparing to the administrative boundary, it is difficult to delineate the shape of urban area, especially the most brightness area (e.g. Zurich and Geneva). On the other hand, the variances between each class cannot reflect the inherent pattern of the spatial data. Jiang et al.(2015) began to extract urban clusters from NTL data in 1992, 2001 and 2010 at the world level, which aim to examine and verify Zip'f law in the world. Jiang(2016) extracted urban clusters from NTL data in China from 1992 to 2012 in order to track China urbanization.

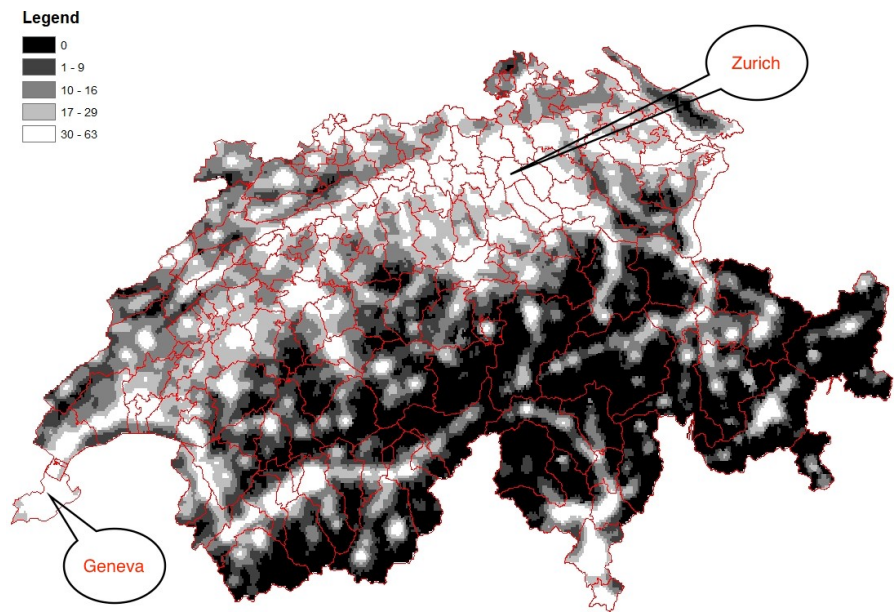

**Figure 1.** Extraction of urban areas uses quantile and overlay on administrative boundary.

## *Head/tail Breaks Classification*

Some of the natural and human phenomena cannot be represented by the normal distribution, such as the gross domestic products (GDP) of different countries. The heavy-tailed distribution is more widespread than the normal distribution in nature, human society and our daily lives. As we know, some low-frequency events, as a long tail in the dataset, play the significant role in our daily life. Figure 2 is a typical example of heavy-tailed distribution, which presents digital number (DN) value of NTL image of Switzerland in 2013. The DN values of pixels are from 0 to 63, among which a large number of pixels is meaningless in 0 and a few of pixels has high DN value that reflects human activities. Apparently, the NTL data also conform to the heavy-tailed distribution. Hence, we need a method to extract the low frequency but important events from a big dataset. However, most of the data classification methods, such as equal interval and natural breaks, focus on high-frequency events. Jiang(2013) introduced and developed a new reverse classification approach to classify data follow the heavy-tailed distribution and pay attention to the low frequency events, which is called head/tail breaks classification.

The head/tail breaks classification was described as "given a variable x, if its value x follows a heavy-tailed distribution, then the mean (m) of the values can divide all the values into two pares: a high percentage in the tail, and a low percentage in the head" (Jiang and Liu 2012). This process should be repeated until the head part values are no longer heavy/tailed distribution. The class intervals are iteratively derived from the arithmetic means. Apparently, the number of classes and the class intervals are naturally determined by the scaling hierarchy of the dataset. Scale is an important and fundamental concept in the geographic domain. Hence, the head/tail division would be beneficial to process spatial data in scaling.

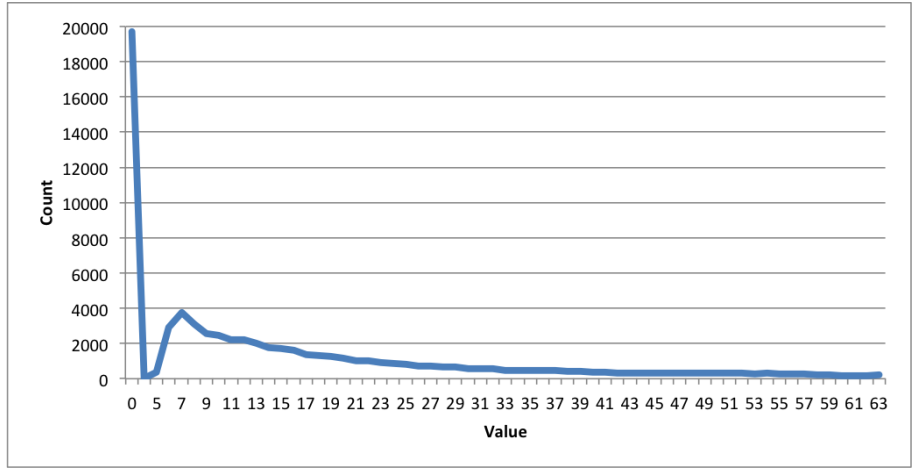

**Figure 2.** A heavy-tailed distribution based on the DN value of NTL image in 2013

Jiang(2013) uses two case studies to compare the nature breaks classification and head/tail breaks classification in different datasets. The results of two case studies demonstrate that the ability of the head/tail breaks classification to capture the inherent hierarchy of the data is better than the nature breaks classification. According to head/tail breaks classification, Jiang and Jia(2011) extracted about 30,000 urban clusters from NTL image in the whole world. Then, the results reflect Zipf's law holds remarkable well for all urban clusters at the global level. Jiang(2016) visualize the city structure and dynamics based on the head/tail breaks classification. Specifically, urban clusters of China, which are extracted from NTL images, are used to track China urbanization in the past two decades. Five largest and rapidly developing regions of China are presented from 1992 to 2012. In general, the NTL imagery data is sorted during a iterative process by the head/tail breaks classification. The head/tail breaks classification is beneficial to classify spatial data that is related to pattern behind the values of data and useful to extract the urban clusters from the spatial data.

## Extraction of Urban Clusters

### Study Area and Data Preparation

The study area for this article relates to all of Switzerland. This work uses two kinds of dataset. One is nighttime light data from 1992 to 2013, which is collected by the US Air Force Defence Meteorological Satellites Program/Operational Lightscan System (DMSP/OLS) allows the monitoring of urban change at regional urban scales. The system was officially acknowledged and declassified in 1972 and shortly after this the low-light detection of cities, fires, fishing boats and gas flares (Elvidge et al.2001). The Operational Linescan System (OLS) is an oscillating scan radiometer designed for cloud imaging with two spectral bands: visiblenear infrared (VNIR, 0.4 1.1m) and thermal infrared (TIR, 10.5 12.6m). The OLS is an oscillating scan radiometer with a broad field of view (3,000km swath) and captures images at a nominal resolution of 0.56km, which is smoothed on-board into 5x5 pixel blocks to 2.8km *. It makes a night- time pass typically between 20.30 and 21.30 each night. Orbiting the Earth 14 times a day means that global coverage can be obtained every 24 hours. The other is street network data of Switzerland, which is swissTLM3D 1.x GDB database in CH1903 LV03 projected coordinate system.

## Urban Clusters Extracted from Nighttime Light Data

This work uses three main steps to extract urban clusters from NTL data (Figure3). The first step is the extraction of study area. According to the administrative boundary of Switzerland, the study area is extracted from the whole world. In order to improve the comparability of NTL data in a time series, the second step is to find the coefficients for the intercalibration of NTL data. The third step is to classify NTL data depend on DN value and extract urban clusters.

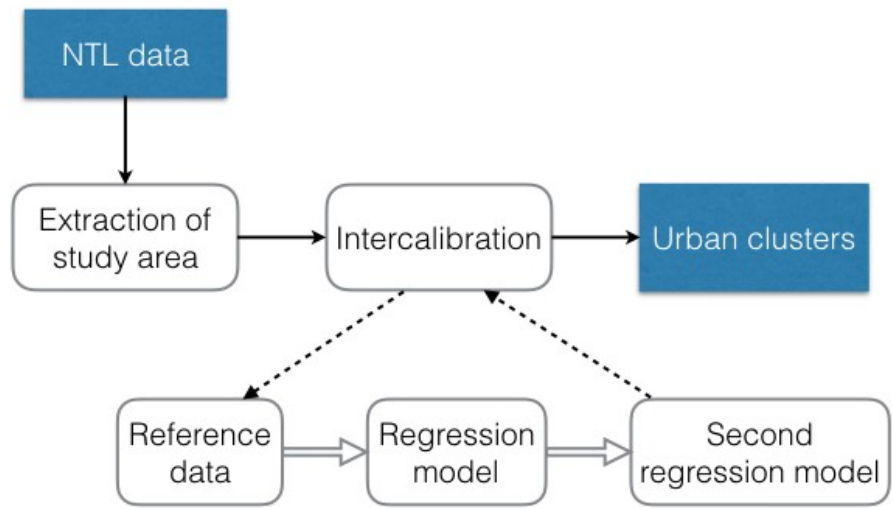

**Figure 3.** Extraction process from NTL data.

---

*http://ngdc.noaa.gov/eog/sensors/ols.html

***Intercalibration of Nighttime Light Data*** A time series of NTL data could capture the dynamics of urbanization despite possible errors from sensor differences. During 1992 to 2013, Version 4 NTL data used six satellites: F10, F12, F14, F15, F16 and F18. A line graph (Figure 4) shows sum of lights in Switzerland from 1992 to 2013. According to the overlay of F15 and F16 from 2004 to 2007, we can find errors of different sensors in a time series of NTL data. Therefore, intercalibration of these data should be performed to improve comparability of NTL data in a time series (Elvidge et al.2009). Liu et al.(2012) illustrated four negative consequences and main problems of NTL imagery data without calibration. First, even if the NTL data in the same year, the DN values derived from different satellites are different. For example, we can find the DN values of F14 and F15 are different in 2000 in Figure 4. Second, different satellites have abnormal fluctuations of the DN values in the same year. Third, the total numbers of lit pixels differed between two satellites in the same year. Fourth, the number of lit pixels derived from the same satellite decreased abnormally between different years.

In order to improve the continuity and comparability of NTL data, this work follows Liu et al.(2012)'s method to correct DN values from 1992 to 2013. First, after compared DN values in the study area from 1992 to 2013, we can find the sum of lights of 2010 is the maximum number in the period 1992 to 2013 (Figure 4). Hence, the NTL data of satellite F18 in 2010 should be used as reference data for intercalibration of other images. Second, a second-order regression (Equation 1) is used to calculate the coefficients for the intercalibration, where y is the dependent variable as the statistical data of the assigned points from the F18 in 2010 which is the reference data, x is the independent variable of the calibrated image and the a, b and c are the coefficients of intercalibration. Third, the regression model (Equation 2) is used to intercalibrated new DN value of other images through the original DN value and the coefficients.

$$y = a + b \times x + c \times x^2, \qquad (1)$$

$$DN_{calibrated} = C_0 + C_1 \times DN + C_2 DN^2 \qquad (2)$$

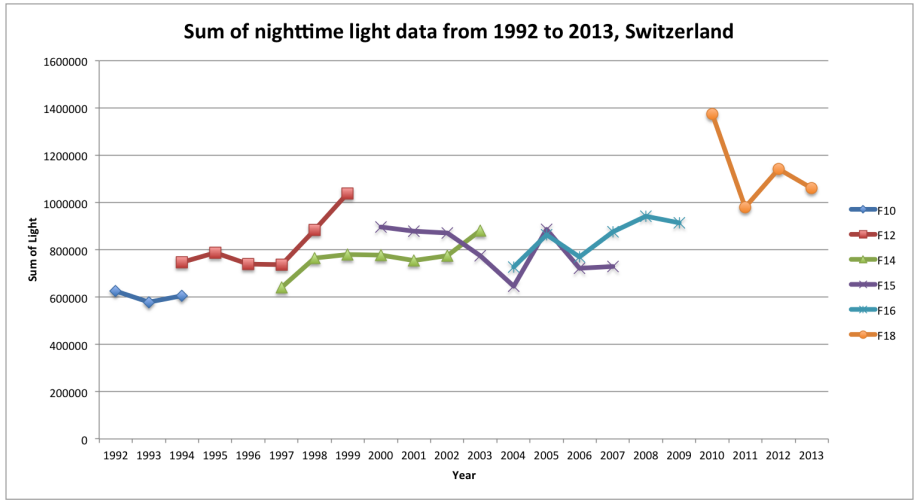

**Figure 4.** Sum of NTL data of Switzerland from 1992 to 2013.

***Classification of DN values*** Because the detected area of the OLS sensors are larger than the spatial geographic extents of the actually associated areas, the error of the NTL data leads to the incorrect indication of the lighted pixels which should not be included in the lighted areas or the source of the night light. In order to reduce the affection of error and identify the effective urban areas for the research, the suitable classification of DN value should be determinate. The head/tail breaks classification only depends on the natural condition of the data. Therefore, this method is beneficial to extract the appropriate urban clusters.

The first step is to find the mean value of DN values as the threshold value and reclassify the pixels of the NTL data. The NTL data of Switzerland contains a total of 70,891 pixels, with DN values between 0 to 63. The first mean value splits all pixels into two unbalance parts. One part includes the pixels that are greater than and equal to the first mean value, which is called Head. The other part includes the pixels that are smaller than the first mean value, which is the Tail. Then, the second mean value of the data values in the head cuts this head into two new head and tail parts. This step should be repeated until the proportion of the final Head part infinitely approaches the 50% but is not larger than 50%. Table 1 shows an example of the calculate the threshold intercalibrated NTL data in 2013. Because the head part is larger than 50%, while the mean is 54.59 at the last row, the classification should be stopped. Because we need to improve the comparability of all the NTL data in the period 1992 to 2013, the final mean values should be calculated based on the mean values each year. After the calculation, three final mean values are used as the candidate threshold values for all the intercalibrated NTL data, which are 19, 34 and 47. This procedure in which the calculation of the mean value in the head/tail breaks classification is necessary to find an optimal threshold value to extract urban clusters.

**Table 1.** Statistics for the head/tail breaks for the Switzerland NTL imagery data in 2009.

| Light | Count | Light*Count | Mean | In head# | In head% | In tail# | In tail% |
|---|---|---|---|---|---|---|---|
| 1-63 | 70891 | 1347627 | 19.0 | 29423 | 41.50% | 41468 | 58.50% |
| 20-63 | 29423 | 1039959 | 35.35 | 12403 | 42.15% | 17020 | 57.85% |
| 36-63 | 12403 | 593531 | 47.85 | 6156 | 49.63% | 6247 | 50.37% |
| 48-63 | 6156 | 336034 | 54.59 | 3287 | 53.4% | 2869 | 46.6% |

At last, the area of urban clusters should be calculated for further analysis. Hence, the raster data need to be vectorized. In this work, ArcGIS 10.2 † is used to realise the vectorization. It provides three methods of intersection solution: geometric, median and none. In order to preserve the angles and straight lines of the features, this work uses the geometric method. Additionally, in order to protect the original shape and reduce the error of area calculation, smoothing weight setting cannot be too large. Figure 5 shows an example of vectorization of NTL data. After the vectorization, all connected pixels lighter than the threshold value are grouped as individual urban clusters. Then, the area of urban clusters can be computed.

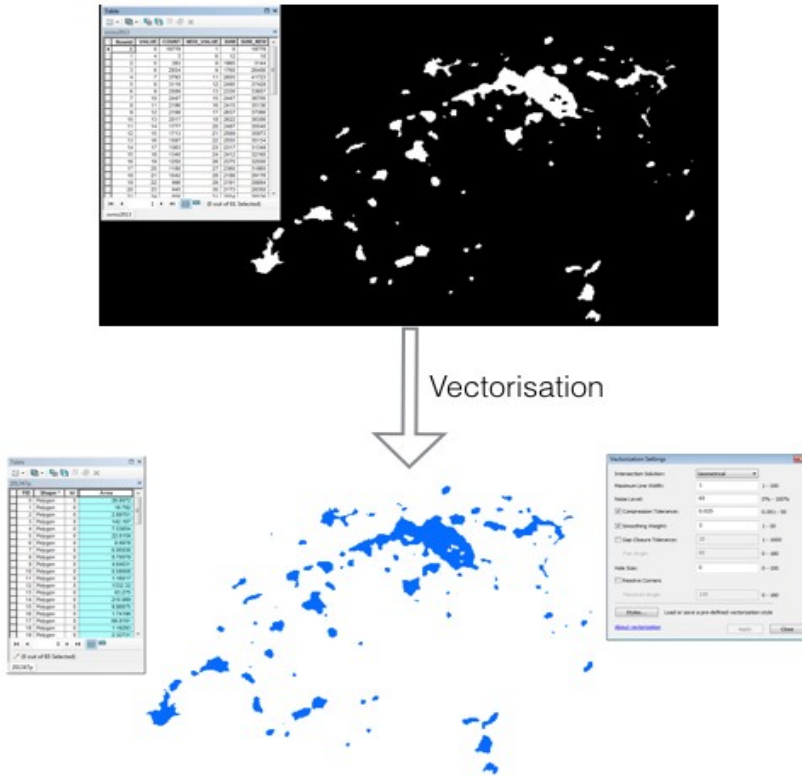

**Figure 5.** Vectorisation of NTL data.

---

†http://help.arcgis.com/en/arcgisdesktop/10.0/help/index.html#/About ArcScan vectorization settings/000w0000000n000000/

## Urban Clusters Extracted from Street Network Data

The urban clusters from street network data can represent human settlement directly and provide a understanding of transportation system development. In this work, urban clusters are the collective of street blocks, which are composed of a set of adjacent street segments (Jiang2015). Hence, the urban clusters are also defined from bottom up. Based on the topological relationship of streets, all the street blocks with inter- relationship constitute the urban clusters. Figure 6 illustrates the extraction process from street network data. This process includes two steps, one is to transform data from geometry to topology, the other is to classify the polygons in order to extract urban clusters.

The original data has 1,586,292 street segments in total (Figure 7a), which includes streets, highway and so on. Because the pattern of street nodes reflects a great extent that of human settlements (Jiang and Jia2011), the street nodes should be derived from street lines during the transformation of data from geometry to topology. Street nodes refer to the sum up of both street junctions or intersections and the ends of streets.

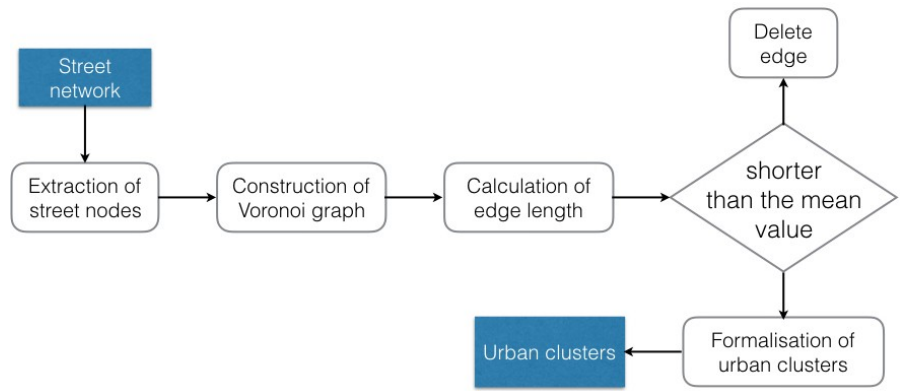

**Figure 6.** Extraction process from street network data.

This work extracts 1,257,219 street nodes (Figure 7b) from original street lines. All the street nodes construct the Voronoi graph (Figure 7c). When the length of Voronoi edge is shorter than the mean value, it should be selected and formalize the urban clusters. From Figure 7c, it is easily found the short edges are far more than long edges. In total, 18,446 polygons (Figure 7d) are extracted from selected Voronoi edges.

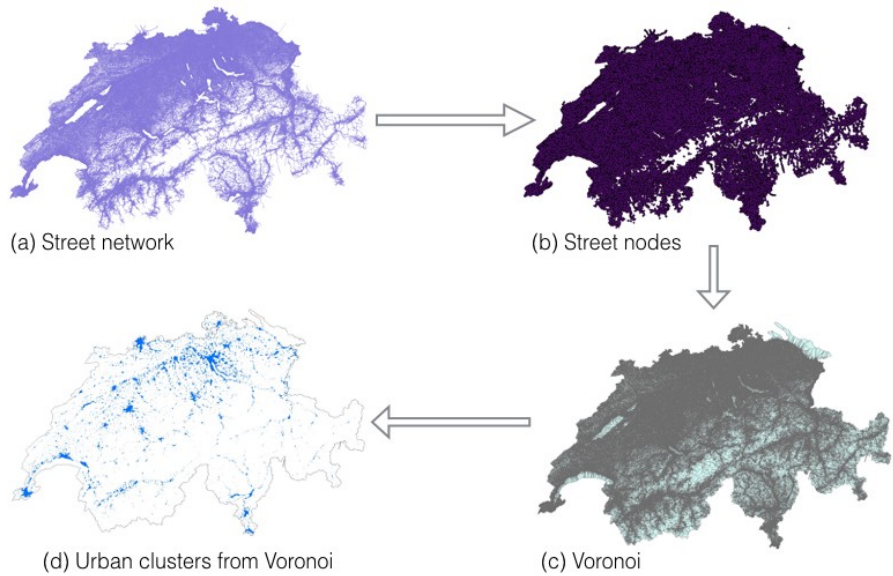

**Figure 7.** The process of urban clusters extraction from street line to polygons.

The connected polygons are generalized as a urban cluster. We also need to determine an optimal threshold value to extract urban clusters from all the polygons. The area of each urban cluster is used to determine the threshold value and classify the urban clusters. The head/tail breaks classification is also used to sort out urban clusters in different groups. Table 2 shows the statistics for the area of urban clusters from Voronoi graph, in which each level has a large tail.

**Table 2.** Statistics for the head/tail breaks for the area of urban clusters from Voronoi graph in Switzerland.

| Area ($km^2$) | Count | Mean | In head# | In head% | In tail# | In tail% |
|---|---|---|---|---|---|---|
| 0-141.182 | 18446 | 0.124 | 1879 | 10% | 16567 | 90% |
| 0.125-141.182 | 1879 | 1.073 | 334 | 18% | 1545 | 82% |
| 1.074-141.182 | 334 | 4.391 | 64 | 19% | 270 | 81% |
| 4.392-141.182 | 64 | 15.002 | 17 | 27% | 47 | 73% |
| 15.003-141.182 | 17 | 35.91 | 5 | 29% | 12 | 71% |

## Results and Discussion

### Results of NTL images

As we mentioned before, the urban clusters are different from real cities, which reflect human settlement and the high aggregation of human activities directly. And urban clusters are derived by the head/tail breaks classification, which means there were no human subjective influences in the acquisition of the cities.

According to the classification of DN values with threshold values (table1), it is necessary to decide which threshold value is better to explain urban clusters in Switzerland. Figure 8 represents a comparison of reclassified results from three candidate threshold value in the Zurich region, among which red lines delineate urban boundary and background is the original NTL data. It is clear find the result from threshold value 19 (Figure 8a) cannot match the brightest areas, while the results from threshold value 34 (Figure 8b) and 47 (Figure 8c) are much better to delineate the boundary of brightest areas. In addition, figure 9 compares the power law distribution of urban clusters from three different threshold values. If the resulting p-value is greater than 0.1 the power law is a plausible hypothesis for the data, otherwise it is rejected. Obviously, the result of threshold value 47 is not fit to power law distribution. In summary, the optimal threshold value is 34.

Compared to the real city with the administrative boundary, some parts are connected as an urban cluster with unclear boundary, especially Zurich region. Hence, the number of urban clusters cannot reflect the growth of urban areas. For instance, the numbers of urban clusters are 91 and 82 in 1992 and 2010 respectively. But in the Figure 10a and 10c, we can easily find the urban clusters' area of 2010 is larger than that of 1992. In addition, we can find the main development of Switzerland focuses on Zurich region. In 1992, Zurich region is composed of several parts. After that, Zurich region is almost connected as a urban cluster and even connected to Basel and Aargau region in 2010. While the urban clusters' area of Zurich region is decreased in 2013 (Figure 10d).

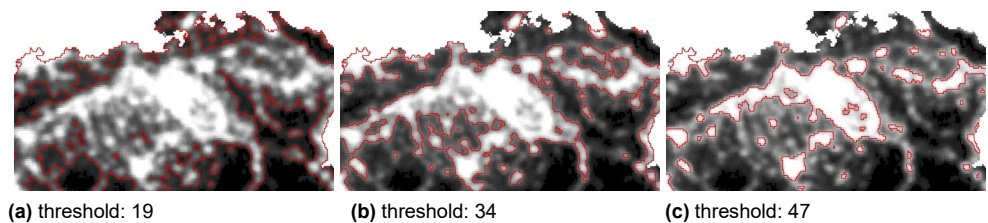

(a) threshold: 19    (b) threshold: 34    (c) threshold: 47

**Figure 8.** The reclassified results (red lines) overlay on the original NTL data in 2013.

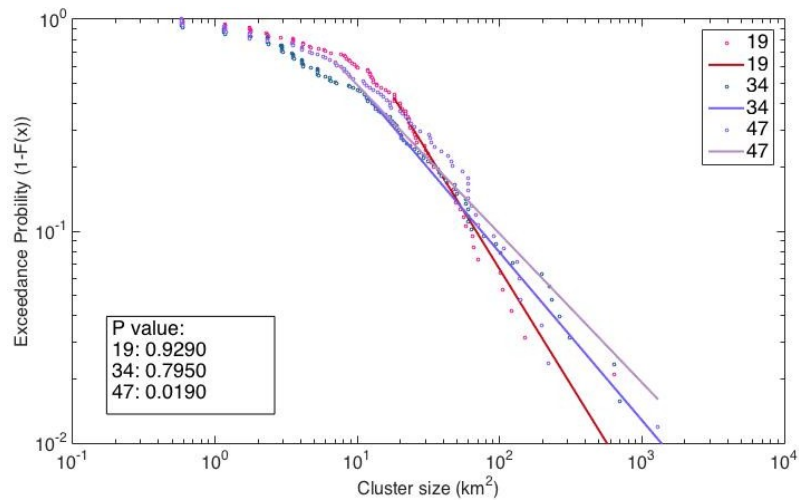

**Figure 9.** Power law distribution of results from three different threshold values.

## *Results of Street Network Data*

Using head/tail break classification, the area of urban clusters is classified in several levels as well (Table 2). In order to determine the optimal threshold value, this work observes two factors of results, power law distribution and the spatial geometric shape of urban clusters. Table 3 lists the total area and amount of urban clusters in each threshold value. Figure 11 shows power law distribution of results from different threshold values. In general, all the results follow power law distribution. But we can find the results of 0.124 $km^2$ are not fit to power law distribution very well.

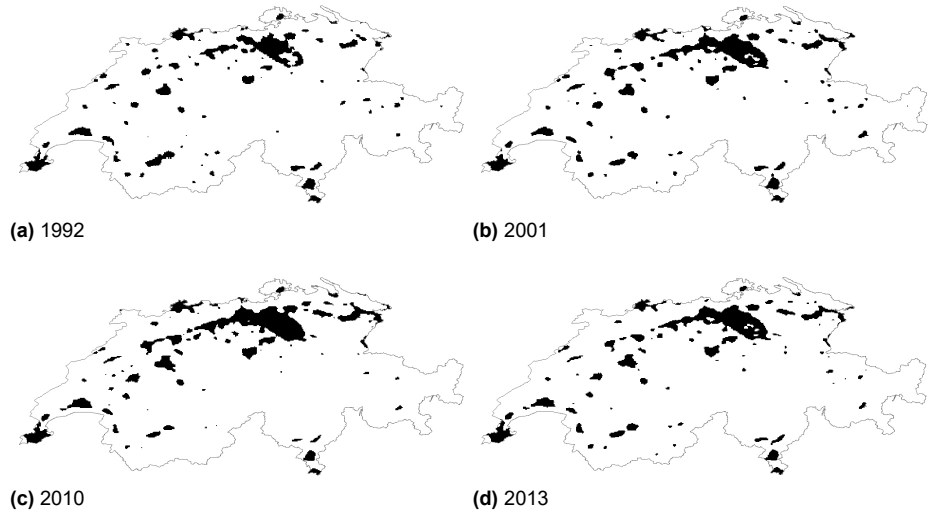

(a) 1992    (b) 2001

(c) 2010    (d) 2013

**Figure 10.** The evolution of urban clusters in Switzerland from 1992 to 2013.

**Table 3.** Statistics for the area and amount of urban clusters in three different threshold values.

| Threshold ($km^2$) | Total area ($km^2$) | Amount |
|---|---|---|
| 0.124 | 2017.75621 | 1879 |
| 1.073 | 1466.75917 | 334 |
| 4.391 | 960.155935 | 64 |

Then, we need to compare the spatial geometric shape of urban clusters from different threshold values. Figure 12 shows results from Voronoi graph in three different threshold value. We can find much more small urban clusters in the Figure 12a. Obviously, this result is not enough to identify the structure of the city at the country level. In addition, the amounts of urban clusters in figure 12c are too small and miss some information at the country level.

Relatively, the shape of urban clusters with threshold value 1.074 *km²* (Figure 12b) is much better than that of others. To sum up the analysis of power law distribution and spatial geometric shape, 1.074 *km²* are selected as threshold to extract urban clusters.

The urban clusters from street network data are classed in four groups by the area. All the urban clusters have irregular shapes. In Figure 13, the biggest regions are Zurich and Basel region as well. In addition, the shapes of urban clusters from street network data are similar with that of real cities.

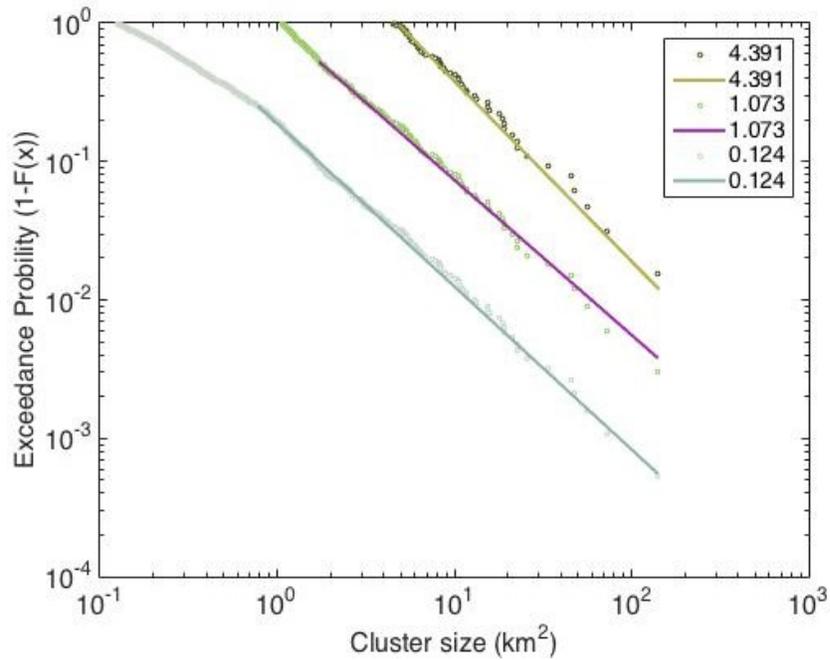

**Figure 11.** Power law distribution of urban clusters.

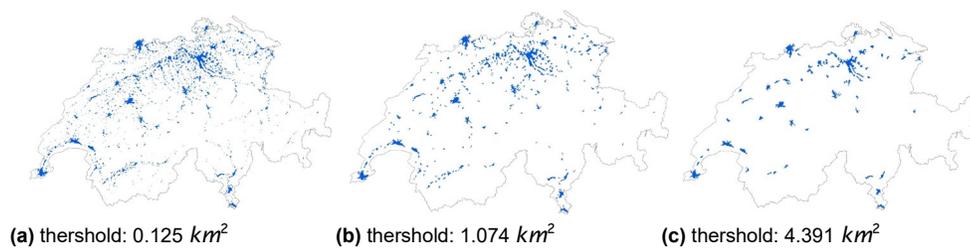

(a) thershold: 0.125 *km²*    (b) thershold: 1.074 *km²*    (c) thershold: 4.391 *km²*

**Figure 12.** Comparison of urban clusters with different mean value from Voronoi graph.

## Comparison and Discussion

Because the lack of street network data in time series, the comparison only relies on the NTL and street network data in 2013. In order to evaluate the results, the urban area of CORINE

Land Cover (CLC) 2012 version 18 is used to compare with urban clusters from geospatial data. CLC was specified to standardize data collection on land in Europe. The urban area of Switzerland from CLC is 2395.156 $km^2$, which is about 6% of the total area. The total area of urban clusters from NTL data is 2833.96 $km^2$, which is about 7% of the total area.

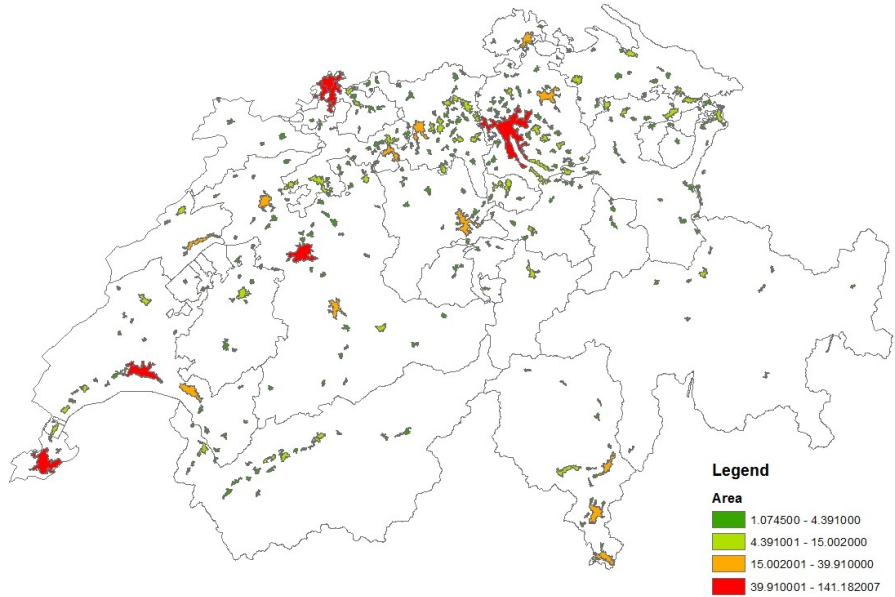

**Figure 13.** Urban clusters from street network data.

The total area of urban clusters from street network data is 1466.759 $km^2$, which is about 4% of the total area. Generally, the percentage of urban clusters from NTL data is closed to that of CLC. When we overlay the urban clusters from geospatial data on the urban area of CLC (Figure 14), both of urban clusters from NTL and street network data are more concentrated than the urban area from CLC. Even though the area of urban clusters from NTL data is similar with that of urban area from CLC, the concentrated urban clusters from NTL data are larger than the others. The shapes of the urban clusters from geospatial data are better to analyse the spatial patter of urban area.

In order to compare results from NTL and street network data, the results have been zooming in to the biggest urban cluster and overlay on the main railway lines on the map. All the results have the similar general geometric shape at the country level. But the shapes of urban clusters from street data are much clearer than that of NTL data. The results from street data only reflect human settlement related to streets. And the results from NTL data include much more information about human behaviors. When we compare to the size of urban clusters, it is clearly find the urban clusters' size from NTL data is larger than that from street data. The largest Swiss urban cluster is located along the "Haerkingen-Olten-Baden-Zurich-Rapperswi" bow, covering an area of about 130 $km$❷ (Figure15a). The street network-based extraction resulted in a network of urban cluster for the same areas (15b), covering less than half of the NTL area. Meanwhile, most of urban clusters are located on the railway lines, especially the junction of railway lines.

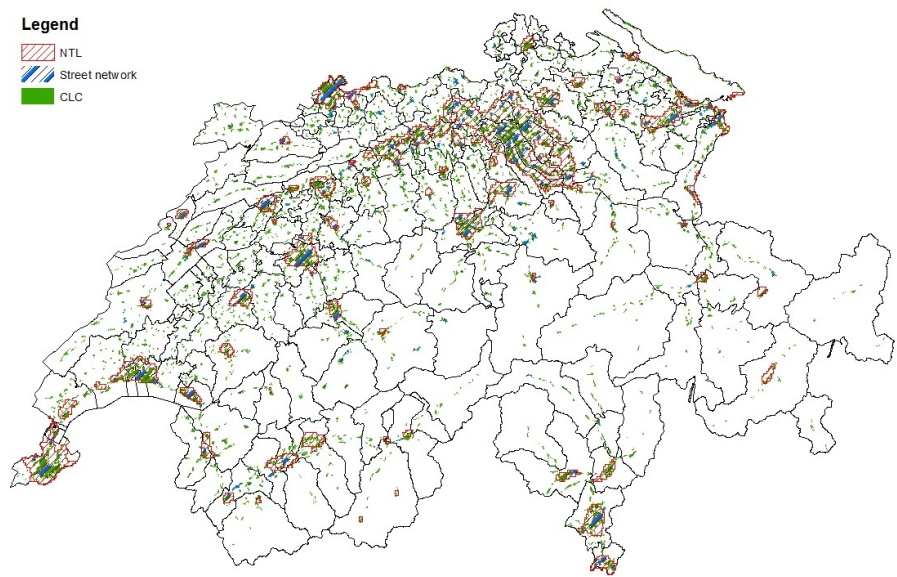

**Figure 14.** Overlay urban clusters on the urban area of CLC.

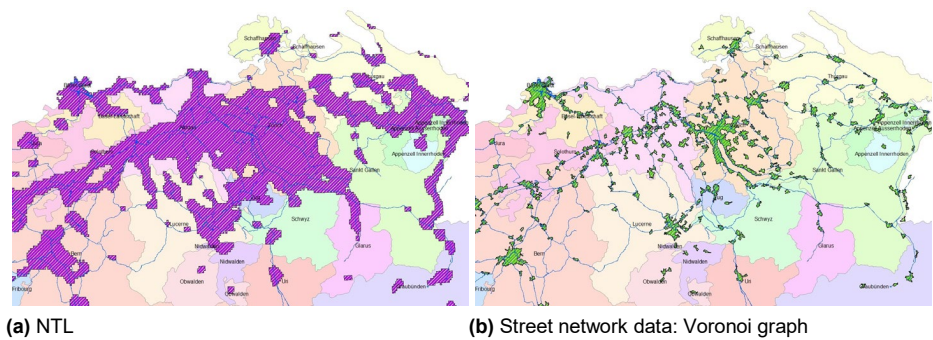

(a) NTL  (b) Street network data: Voronoi graph

**Figure 15.** Overlay urban clusters on administrative districts and railway (blue lines).

## Conclusion and Further Works

The purpose of this work is to identify urban area from geospatial data that can reflect human activities directly. The concept of urban clusters provides a powerful insight to capture data and observe urban area. This work uses head/tail breaks classification to extract urban clusters from NTL imagery data in Switzerland during 1992 to 2013 and street data in 2013. Compared to the real city with administrative boundary, the results of this work have no subjective views of human influence. From all the results of urban clusters, we can find the growth of the urban area is nonlinear in Switzerland. In general, all the urban clusters follow the power law distribution at the country level. It can reflect their neighbors can influence the evolution of small regions. Hence, when we research on the small cities or countries, it is better to consider their neighbors as well.

The investigation proved the feasibility of the 2 extraction approaches, but also the significance difference, which is not acceptable for delineation purposes. Since the threshold values for the classification of "urban" entities had to be defined by the analyst, an approach should be developed that brings the 2 classification methods together, thus overcoming the sensitivity to manual threshold values. Further work is better to include historical data of street to learn the growth of cities deeply. In order to derive much more information at the city level, socioeconomic variables (e.g. GPD and energy use) should be integrated to understand the growth and evolution of city. Meanwhile, the NTL data always involves a lot of different information about geographic features. Further work also needs to improve the extraction of valid or necessary information from the NTL data.

## Acknowledgements

The research was supported by COST TU1305. The author would like to thank Dr. Bin Jiang for his comments and suggestion.

The research was conducted at the Future Resilient Systems at the Singapore-ETH Centre, which was established collaboratively between ETH Zurich and Singapores National Research Foundation (FI 370074011) under its Campus for Research Excellence and Technological Enterprise programme.